\documentclass[a4paper,12pt]{article}
\usepackage{multicol}
\usepackage{latexsym}
\usepackage{graphics}
\usepackage{amssymb}
\usepackage{a4wide}
\usepackage{amsmath}

\begin{document}
\title{Geometric Numerical Integration Applied to  \\
  The Elastic Pendulum at Higher Order Resonance }
\author{
\begin{tabular}{cc}
  J. M. Tuwankotta\footnote{On leave from Jurusan Matematika, FMIPA,Institut Teknologi Bandung,
  Ganesha no. 10, Bandung, Indonesia}\;\; \footnote{Corresponding author}   & G.R.W. Quispel  \\
  \small E-mail: tuwankotta@math.uu.nl & \small Email: R.Quispel@latrobe.edu.au \\
  \small Mathematisch Instituut, Universiteit
Utrecht    &\small School of Mathematical and Statistical
Sciences\\ \small PO Box 80.010, 3508 TA Utrecht   & \small La
Trobe University, Bundoora, Vic. 3083 \\\small  The Netherlands
&\small  Australia \end{tabular}}

\maketitle

\begin{abstract}
In this paper we study the performance of a symplectic numerical
integrator based on the splitting method.  This method is applied
to a subtle problem i.e. higher order resonance of the elastic
pendulum.  In order to numerically study the phase space of the
elastic pendulum at higher order resonance, a numerical integrator
which preserves qualitative features after long integration times
is needed.  We show by means of an example that our symplectic
method offers a relatively cheap and accurate numerical
integrator.
\end{abstract}

\vskip -0.25 cm
{\small \noindent  {\bf Keywords.} Hamiltonian mechanics,
higher-order resonance, elastic pendulum, symplectic numerical
integration, geometric integration.}

\section{Introduction}

Higher order resonances are known to have a long time-scale
behaviour.  From an asymptotic point of view, a first order
approximation (such as first order averaging) would not be able to
clarify the interesting dynamics in such a system. Numerically,
this means that the integration times needed to capture such
behaviour are significantly increased.  In this paper we present a
reasonably cheap method to achieve a qualitatively good result
even after long integration times.

Geometric numerical integration methods for (ordinary)
differential equations (\cite{ Budd,MQ, calvo}) have emerged in
the last decade as alternatives to traditional methods (e.g.
Runge-Kutta methods). Geometric methods are designed to preserve
certain properties of a given ODE exactly (i.e. without truncation
error). The use of geometric methods is particularly important for
long integration times.  Examples of geometric integration methods
include symplectic integrators, volume-preserving integrators,
symmetry-preserving integrators, integrators that preserve first
integrals (e.g. energy), Lie-group integrators, etc.  A survey is
given in \cite{MQ}.

It is well known that resonances play an important role in
determining the dynamics of a given system.  In practice, higher
order resonances occur more often than lower order ones, but their
analysis is more complicated.  In \cite{Sander}, Sanders was the
first to give an upper bound on the size of the resonance domain
(the region where interesting dynamics takes place) in two degrees
of freedom Hamiltonian systems.  Numerical studies by van den
Broek \cite{vdbroek}, however, provided evidence that the
resonance domain is actually much smaller.  In \cite{tv},
Tuwankotta and Verhulst derived improved estimates for the size of
the resonance domain, and provided numerical evidence that for the
$4:1$ and the $6:1$ resonances of the elastic pendulum, their estimates are sharp.  The
numerical method they used in their analysis \footnote{A
Runge-Kutta method of order $7-8$}, however, was not powerful
enough to be applied to higher order resonances.  In this paper we
construct a symplectic integration method, and use it to show
numerically that the estimates of the size of the resonance domain
in \cite{tv} are also sharp for the $4:3$ and the $3:1$
resonances.  

Another subtle problem regarding to this resonance manifold is the bifurcation of this manifold as the energy increases.  To study this problem numerically one would need a numerical method which is reasonably cheap and accurate after a long integration times.    

In this paper we will use the elastic pendulum as an example.  The
elastic pendulum is a well known (classical) mechanical problem
which has been studied by many authors.  One of the reasons is
that the elastic pendulum can serve as a model for many problems
in different fields.  See the references in \cite{georgiou,tv}. In
itself, the elastic pendulum is a very rich dynamical system. For
different resonances, it can serve as an example of a chaotic
system, an auto-parametric excitation system (\cite{vdburg1}), or
even a linearizable system.  The system also has (discrete) symmetries
which turn out to cause degeneracy in the normal form.

We will first give a brief introduction to the splitting method
which is the main ingredient for the symplectic integrator in this
paper.  We will then collect the analytical results on the elastic
pendulum that have been found by various authors.  Mostly, in this
paper we will be concerned with the higher order resonances in the
system. All of this will be done in the next two sections of the
paper. In the fourth section we will compare our symplectic
integrator with the standard 4-th order Runge-Kutta method and
also with an order $7-8$ Runge-Kutta method.  We end the fourth
section by calculating the size of the resonance domain of the
elastic pendulum at higher order resonance.


\section{Symplectic Integration \label{Symplectic}}
Consider a symplectic space $\Omega = \mathbb{R}^{2n}, n \in
\mathbb{N}$ where each element $\boldsymbol{\xi}$ in $\Omega$ has
coordinate $(\boldsymbol{q},\boldsymbol{p})$ and the symplectic
form is $\text{d}\boldsymbol{q}\wedge \text{d}\boldsymbol{p}$.
For any two functions $F,G \in \mathcal{C}^{\infty}(\Omega)$
define
\begin{equation*}
   \{F,G\} = \sum \limits_1^n \left(\frac{\partial F}{\partial q_j}
      \frac{\partial G}{\partial p_j} - \frac{\partial G}{\partial q_j}
           \frac{\partial F}{\partial p_j} \right) \in \mathcal{C}^{\infty}(\Omega),
\end{equation*}  which is called the Poisson bracket of $F$ and $G$.  Every function
 $H \in \mathcal{C}^{\infty}(\Omega)$ generates a (Hamiltonian) vector field defined by
 $\{q_i,H \},\{p_i,H \}, i= 1,\ldots, n $.  The dynamics of $H$ is then governed by the
 equations of motion of the form
\begin{equation*}
  \begin{split}
     \dot{q_i} = & \{q_i,H \} \\
     \dot{p_i} = & \{p_i,H \}, \quad i =1,\ldots, n.
  \end{split}
\label{eqnofmotion}
\end{equation*}  Let $X$ and $Y$ be two Hamiltonian vector fields, defined in $\Omega$,
associated with Hamiltonians $H_{\scriptscriptstyle X}$ and
$H_{\scriptscriptstyle Y}$  in $\mathcal{C}^{\infty}(\Omega)$
respectively.  Consider another vector field $[X,Y]$ which is just
the commutator of the vector fields $X$ and $Y$ . Then $[X,Y]$ is
also a Hamiltonian vector field with Hamiltonian
$H_{\scriptscriptstyle [X,Y]} = \{H_{\scriptscriptstyle
X},H_{\scriptscriptstyle Y}\}$. See for example  \cite{arnold, MR, olver} for details.

We can write the flow of the Hamiltonian vector fields $X$ as
\begin{equation*}
      \varphi_{{\scriptscriptstyle X};t} =  \text{ exp}(tX) \equiv I + tX
      + \frac{1}{2!}(tX)^2 + \frac{1}{3!}(tX)^3 + \cdots  
\end{equation*} (and so does the flow of $Y$).  By the BCH formula, there exists a (formal) Hamiltonian vector
field $Z$ such that
\begin{equation}
   Z =  (X + Y) + \frac{t}{2}[X,Y] + \frac{t^2}{12}\left([X,X,Y] + [Y,Y,X]\right) + O(t^3)
\label{firstorder}
\end{equation} and $\text{exp}(tZ) = \text{exp}(tX)\text{exp}(tY)$, where $[X,X,Y] = [X,[X,Y]]$
, and so on.  Moreover, Yoshida (in \cite{Yoshida}) shows that $\text{exp}(tX)
\text{exp}(tY)\text{exp}(tX) = \text{exp}(tZ),$ where
\begin{equation}
  Z = (2X + Y) + \frac{t^2}{6}\left([Y,Y,X] -[X,X,Y]\right) + O(t^4).
\label{secondorder}
\end{equation}  We note that in terms of the flow, the multiplication of
the exponentials above means composition of the corresponding
flow, i.e. $\varphi_{{\scriptscriptstyle  Y};t} \circ
\varphi_{{\scriptscriptstyle X};t}$.

Let $\tau \in \mathbb{R}$ be a small positive number and consider a Hamiltonian
system with Hamiltonian $H(\boldsymbol{\xi}) = H_{\scriptscriptstyle X}(\boldsymbol{\xi})
 + H_{\scriptscriptstyle Y}(\boldsymbol{\xi})$, where $\boldsymbol{\xi} \in \Omega$, and
 $\dot{\boldsymbol{\xi}} = X + Y$.  Using (\ref{firstorder}) we have that
$\varphi_{{\scriptscriptstyle Y};\tau} \circ
\varphi_{{\scriptscriptstyle X};\tau}$ is (approximately) the flow
of a Hamiltonian system
\begin{equation*}
   \dot{\boldsymbol{\xi}} =  (X + Y) + \frac{\tau}{2}[X,Y] + \frac{\tau^2}{12}\left([X,X,Y]
   + [Y,Y,X]\right) + O(\tau^3),
\end{equation*} with Hamiltonian
\begin{equation*}
    H_{\tau} = H_{\scriptscriptstyle X} + H_{\scriptscriptstyle Y} + \frac{\tau}{2}
    \left\{H_{\scriptscriptstyle X}, H_{\scriptscriptstyle Y}\right\} +
   \frac{\tau^2}{12} \left(\left\{ H_{\scriptscriptstyle X},H_{\scriptscriptstyle X},
    H_{\scriptscriptstyle Y} \right\}  +\left\{ H_{\scriptscriptstyle Y},
    H_{\scriptscriptstyle Y}, H_{\scriptscriptstyle X} \right\}\right) + O(\tau^3).
\end{equation*}  Note that $\{H,K,F\} = \{H,\{K,F\}\}$.
This mean that $H - H_{\tau} = O(\tau)$ or, in other words
\begin{equation}
\varphi_{{\scriptscriptstyle Y};\tau} \circ \varphi_{{\scriptscriptstyle X};\tau}
 = \varphi_{\scriptscriptstyle X+Y}(\tau) +  O(\tau^2).
\label{firstsplit}
\end{equation}  As before and using (\ref{secondorder}), we conclude that
\begin{equation}
  \varphi_{{\scriptscriptstyle X};\frac{\tau}{2}} \circ \varphi_{{\scriptscriptstyle Y};\tau}
  \circ \varphi_{{\scriptscriptstyle X};\frac{\tau}{2}} = \varphi_{\scriptscriptstyle X+Y}(\tau)
   +  O(\tau^3).
\label{secondsplit}
\end{equation}

Suppose that $\psi_{{\scriptscriptstyle X};\tau}$ and
$\psi_{{\scriptscriptstyle Y};\tau}$
 are numerical integrators of system $\dot{\boldsymbol{\xi}} = X$ and
 $\dot{\boldsymbol{\xi}} = Y$ (respectively).  We can use symmetric composition (see \cite{Mc2}) to improve
 the accuracy of $\psi_{{\scriptscriptstyle X+Y};\tau}$.  If  $\psi_{{\scriptscriptstyle Y};
 \tau}$ and $ \psi_{{\scriptscriptstyle X};\tau}$ are symplectic, then the composition forms
 a symplectic numerical integrator for $X+Y$.  See \cite{calvo} for more discussion; also
 \cite{MQ} for references.  If we can split $H$ into two (or more) parts which Poisson commute
 with each other (i.e. the Poisson brackets between each pair vanish), then we have $H=H_\tau$.
   This implies that in this case the accuracy of the approximation depends only on the accuracy of the
   integrators for $X$ and $Y$.  An example of this case is when we are integrating the Birkhoff
   normal form of a Hamiltonian system.

\section{The Elastic Pendulum}
\begin{figure}[h]
  \begin{center}
    \resizebox{.35\textwidth}{!}{
      \includegraphics{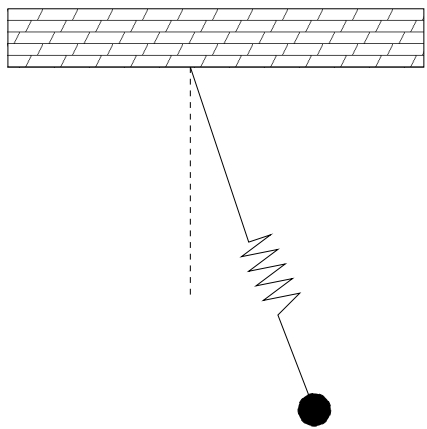}}
  \end{center}
  \caption{
    \label{fig4}
The elastic pendulum.}
\end{figure}

Consider a spring with spring constant $s$ and length $l_\circ$ to
which a mass $m$ is attached.  Let $g$ be the gravitational
constant and $l$ the length of the spring under load in the
vertical position, and let $r$ be the distance between the mass
$m$ and the suspension point.  The spring can both oscillate in
the radial direction and swing like a pendulum. This is called the
{\it elastic pendulum}. See Figure (\ref{fig4}) for illustration
and \cite{tv} (or \cite{vdburg2}) for references.

The phase space is $\mathbb{R}^4$ with canonical coordinate
$\boldsymbol{\xi}= (z,\varphi,p_z,p_\varphi)$, where
$z=(r-l_\circ)/l_\circ$.  Writing the linear frequencies of the
Hamiltonian as $\omega_z = \sqrt{s/m}$ and $\omega_\varphi =
\sqrt{g/l}$, the Hamiltonian of the elastic pendulum becomes
\begin{equation}
   H = \frac{1}{2\sigma} \left( p_z^2 + \frac{p_\varphi^2}{(z+1)^2} \right)
      + \frac{\sigma}{2}{\omega_z}^2 \left( z +
\left(\frac{\omega_\varphi}{\omega_z}\right)^2 \right)^2 - \sigma
{\omega_{\varphi}}^2 (z+1)\cos\varphi, \label{transform}
\end{equation} where $\sigma = ml^2$.  By choosing the right physical
dimensions, we can scale out $\sigma$.  We remark that for the
elastic pendulum as illustrated in Figure \ref{fig4}, we have
$\omega_z \le \omega_\varphi$.  See \cite{tv} for details. It is
clear that this system possesses symmetry

\begin{equation}
  T : (z,\varphi,p_z,p_\varphi,t) \mapsto (z,-\varphi,p_z,-p_\varphi,t)
  \label{Tsymmetry}
\end{equation}
and the reversing symmetries

\begin{equation}
\begin{array}{c}
  R_1 : (z,\varphi,p_z,p_\varphi,t) \mapsto
  (z,\varphi,-p_z,-p_\varphi,-t), \\
  R_2 : (z,\varphi,p_z,p_\varphi,t) \mapsto
  (z,-\varphi,-p_z,p_\varphi,-t).
\end{array}
\label{Rsymmetry}
\end{equation}

If there exist two integers $k_1$ and $k_2$ such that $k_1
\omega_z + k_2 \omega_\varphi = 0$, then we say $\omega_z$ and
$\omega_\varphi$ are in {\it resonance}.  Assuming
$(|k_1|,|k_2|)=1$, we can divide the resonances in two types, e.g.
lower order resonance if $|k_1| + |k_2| < 5$ and higher order
resonance if $|k_1| + |k_2| \ge 5$.  In the theory of normal
forms, the type of normal form of the Hamiltonian is highly
dependent on the type of resonance in the system.  See
\cite{arnold}.

In general, the elastic pendulum has at least one fixed point
which is the origin of phase space.  This fixed point is elliptic.
For some of the resonances, there is also another fixed point
which is of the saddle type, i.e. $(z,\varphi,p_z,p_\varphi) =
(-2(\omega_\varphi/\omega_z)^2,\pi,0,0)$.  From the definition of
$z$, it is clear that the latter fixed point only exists for
$\omega_z/\omega_\varphi > \sqrt{2}$.  The elastic pendulum also
has a special periodic solution in which $\varphi = p_\varphi =0$
(the normal mode). This normal mode is an exact solution of the
system derived from (\ref{transform}).  We note that there is no
nontrivial solution of the form $(0,\varphi(t),0,p_\varphi(t))$.

Now we turn our attention to the neighborhood of the origin. We
refer to \cite{tv} for the complete derivation of the following
Taylor expansion of the Hamiltonian (we have dropped the bar)
\begin{equation}
    H = H_2 + \varepsilon H_3 + \varepsilon^2 H_4 + \varepsilon^3 H_5 + \cdots,
\label{expand}
\end{equation}
with
\begin{equation*}
 \begin{split}
    H_2 =& \textstyle \frac{1}{2}\omega_z \left( z^2 + p_z^2 \right)
       + \frac{1}{2}\omega_\varphi \left(\varphi^2 + p_\varphi^2 \right)  \\
    H_3 =& \textstyle \frac{\omega_\varphi}{\sqrt{ \omega_z}} \left( \frac{1}{2}
         z\varphi^2  -z p_\varphi^2 \right) \\
    H_4 =& \textstyle \left( \frac{3}{2}\frac{\omega_\varphi}{\omega_z} z^2
         p_\varphi^2 - \frac{1}{24}\varphi^4  \right)\\
    H_5 =&\textstyle -\frac{1}{ \sqrt{ \omega_z}} \left(\frac{1}{24} z \varphi^4
          + 2 \frac{\omega_\varphi}{\omega_z}  z^3 p_\varphi^2 \right)  \\
         &  \qquad   \vdots
 \end{split}
\end{equation*}

In \cite{vdburg1} the $2:1$-resonance of the elastic pendulum has
been studied intensively.  At this specific resonance, the system
exhibits an interesting phenomenon called auto-parametric
excitation, e.g. if we start at any initial condition arbitrarily
close to the normal mode, then we will see energy interchanging
between the oscillating and swinging motion. In
\cite{duistermaat}, the author shows that the normal mode solution
(which is the vertical oscillation) is unstable and therefore,
gives an explanation of the auto-parametric behavior.

Next we consider two limiting cases of the resonances, i.e. when
$\omega_z/\omega_\varphi \rightarrow \infty$ and
$\omega_z/\omega_\varphi \rightarrow 1$.  The first limiting case
can be interpreted as a case with a very large spring constant so
that the vertical oscillation can be neglected.  The spring
pendulum then becomes an ordinary pendulum; thus the system is
integrable. The other limiting case is interpreted as the case
where $l_\circ =0$ (or very weak spring) \footnote{This case is
unrealistic for the model illustrated in Figure \ref{fig4}.  A
more realistic mechanical model with the same Hamiltonian
(\ref{transform}) can be constructed by only allowing some part of
the spring to swing}. Using the transformation $r = l(z+1)$, $x =
r\cos\varphi$ and $y = r\sin\varphi,$ we transform the Hamiltonian
(\ref{transform}) to the Hamiltonian of the harmonic oscillator.
Thus this case is also integrable. Furthermore, in this case all
solutions are periodic with the same period which is known as
isochronism. This means that we can remove the dependence of the
period of oscillation of the mathematical pendulum on the
amplitude, using this specific spring. We note that this
isochronism is not derived from the normal form (as in
\cite{vdburg2})  but exact.

All other resonances are higher order resonances. In two degrees
of freedom (which is the case we consider), for fixed small energy
the phase space of the system near the origin looks like the phase space of  decoupled harmonic oscillator.     A consequence of this fact is that in the
neighborhood of the origin, there is no interaction between the
two degrees of freedom.  The normal mode (if it exists), then
becomes elliptic (thus stable).

Another possible feature of this type of resonance is the
existence of a {\it resonance manifold} containing periodic
solutions (see \cite{guckenheimer} paragraph 4.8).  We remark that
the existence of this resonance manifold does not depend on
whether the system is integrable or not. In the {\it resonance
domain} (i.e. the neighborhood of the resonant manifold),
interesting dynamics (in the sense of energy interchanging between
the two degrees of freedom) takes place (see \cite{Sander}). Both
the size of the domain where the dynamics takes place and the
time-scale of interaction are characterized by $\varepsilon$ and
the order of the resonance, i.e. the estimate of the size of the
domain is
\begin{equation}
\displaystyle d_\varepsilon = O\left(\varepsilon^{\frac{m+n
-4}{2}}\right)\end{equation} and the time-scale of interactions is
$O(\varepsilon^{-\frac{m+n}{2}})$ for $\omega_z : \omega_\varphi =
m:n$ with $(m,n)=1$. \footnote{Due to a particular symmetry, some
of the lower order resonances become higher order resonances
(\cite{tv}).  In those cases, $(m,n)=1$ need not hold.} We note
that for some of the higher order resonances where
$\omega_z/\omega_\varphi \approx 1$ the resonance manifold fails
to exist. See \cite{tv} for details.

\section{Numerical Studies on the Elastic Pendulum}
One of the aims of this study is to construct a numerical
Poincar\'e map ($\mathcal{P}$) for the elastic pendulum in higher
order resonance.  As is explained in the previous section,
interesting dynamics of the higher order resonances takes place in
a rather small part of phase space. Moreover, the interaction
time-scale is also rather long. For these two reasons, we need a
numerical method which preserves qualitative behavior after a long
time of integration.  Obviously by decreasing the time step of any
standard integrator (e.g. Runge-Kutta method), we would get a
better result.  As a consequence however, the actual computation
time would become prohibitively long. Under these constraints, we
would like to propose by means of an example that symplectic
integrators offer reliable and reasonably cheap methods to obtain
qualitatively good phase portraits.

We have selected four of the most prominent higher order
resonances in the elastic pendulum. For each of the chosen
resonances, we derive its corresponding equations of motion from
(\ref{expand}). This is done because the dependence on the small
parameter $\varepsilon$ is more visible there than in
(\ref{transform}). Also from the asymptotic analysis point of
view, we know that (\ref{expand}) truncated  to a sufficient
degree has enough ingredients of the dynamics of
(\ref{transform}).

The map $\mathcal{P}$ is constructed as follows. We choose the
initial values $\boldsymbol{\xi_\circ}$ in such a way that they
all lie in the approximate energy manifold $H_2 = E_\circ \in
\mathbb{R}$ and in the section $\Sigma = \{\boldsymbol{\xi} =
(z,\varphi,p_z,p_\varphi)|z=0,p_z>0\}$.  We follow the numerically
constructed trajectory corresponding to $\boldsymbol{\xi_\circ}$
and take the intersection of the trajectory with section $\Sigma$.
The intersection point is defined as
$\mathcal{P}(\boldsymbol{\xi_\circ})$.  Starting from
$\mathcal{P}(\boldsymbol{\xi_\circ})$ as an initial value, we go
on integrating and in the same way we find
$\mathcal{P}^2(\boldsymbol{\xi_\circ})$, and so on.

The best way of measuring the performance of a numerical
integrator is by comparing with an exact solution.  Due to the
presence of the normal mode solution (as an exact solution), we
can check the performance of the numerical integrator in this way
(we will do this in section \ref{sec42}) . Nevertheless, we should
remark that none of the nonlinear terms play a part in this normal
mode solution. Recall that the normal mode is found in the
invariant manifold $\{(z,\varphi, p_z, p_\varphi | \varphi =
p_\varphi = 0 \}$ and in this manifold the equations of motion of
(\ref{expand}) are linear.

Another way of measuring the performance of an integrator is to
compare it with other methods.  One of the best known methods for
time integration are the Runge-Kutta methods (see \cite{Hairer}).
We will compare our integrator with a higher order ($7$-$8$ order)
Runge-Kutta method (RK78). The RK78 is based on the method of
Runge-Kutta-Felbergh (\cite{Stoer}). The advantage of this method
is that it provides step-size control.  As is indicated by the
name of the method, to choose the optimal step size it compares
the discretizations using $7$-th order and $8$-th order
Runge-Kutta methods. A nice discussion on lower order methods of
this type, can be found in \cite{Stoer} pp. 448-454.  The
coefficients in this method are not uniquely determined. For RK78
that we used in this paper, the coefficients were calculated by C.
Simo from the University of Barcelona.  We will also compare the
symplectic integrator (SI) to the standard $4$-th order
Runge-Kutta method.

We will first describe the splitting of the Hamiltonian which is
at the core of the symplectic integration method in this paper. By
combining the flow of each part of the Hamiltonian, we construct a
$4$-th order symplectic integrator.  The symplecticity is obvious
since it is the composition of exact Hamiltonian flows.  Next we
will show the numerical comparison between the three integrators,
RK78, SI and RK4.  We compare them to an exact solution. We will
also show the performance of the numerical integrators with
respect to energy preservation. We note that SI are not designed
to preserve energy (see \cite{MQ}).  Because RK78 is a higher
order method (thus more accurate), we will also compare the orbit
of RK4 and SI.  We will end this section with results on the size
of the resonance domain calculated by the SI method.

\subsection{The Splitting of the Hamiltonian \label{sec41}}
Consider again the expanded Hamiltonian of the elastic pendulum
(\ref{expand}). We split this Hamiltonian into integrable parts:
$H = H^1 + H^2 + H^3$, where
\begin{equation}
 \begin{split}
    H^1 = &  \varepsilon \frac{\omega_\varphi}{2\sqrt{ \omega_z}} z\varphi^2
            - \varepsilon^2 \frac{1}{24}\varphi^4 -
              \varepsilon^3 \frac{1}{24 \sqrt{ \omega_z}} z \varphi^4 + \cdots \\
    H^2 = &  - \varepsilon  \frac{\omega_\varphi}{\sqrt{ \omega_z}} z p_\varphi^2
            + \varepsilon^2 \frac{3}{2}\frac{\omega_\varphi}{\omega_z} z^2 p_\varphi^2
             - \varepsilon^3 \frac{2 \omega_\varphi}{\omega_z\sqrt{ \omega_z}}  z^3 p_\varphi^2 + \cdots \\
    H^3 =& \textstyle \frac{1}{2}\omega_z \left( z^2 + p_z^2 \right)
            + \frac{1}{2}\omega_\varphi \left(\varphi^2 + p_\varphi^2 \right).
 \end{split}
\label{splitting}
\end{equation}  Note that the equations of motion derived from each part of the
Hamiltonian can be integrated exactly; thus we know the exact flow
$\varphi_{1;\tau}$, $\varphi_{2:\tau}$, and $\varphi_{3;\tau}$
corresponding to $H^1$,$H^2$, and $H^3$ respectively. This
splitting has the following advantages.
\begin{itemize}
\item It preserves the Hamiltonian structure of the system.
\item It preserves the symmetry (\ref{Tsymmetry}) and reversing symmetries
(\ref{Rsymmetry}) of $H$.
\item $H^1$ and $H^2$ are of $O(\varepsilon)$ compared with $H$ (or $H^3$).
\end{itemize}
Note that, for each resonance we will truncate (\ref{splitting}) up to and including the degree where the resonant terms  of the lowest order occur.

We define

\begin{equation}
\varphi_{\tau} = \varphi_{1;\tau/2}\circ \varphi_{2;\tau/2}\circ
\varphi_{3;\tau} \circ \varphi_{2;\tau/2} \circ
\varphi_{1;\tau/2}. \end{equation}  From section \ref{Symplectic}
we know that this is a second order method.  Next we define
$\gamma = 1/(2 - \sqrt[3]{2})$ and $\psi_\tau = \varphi_{\gamma
\tau} \circ \varphi_{(1-2\gamma) \tau} \circ \varphi_{\gamma
\tau}$ to get a fourth order method.  This is known as the
generalized Yoshida method (see \cite{MQ}). By, Symplectic
Integrator (SI) we will mean this fourth order method.  This
composition preserves the symplectic structure of the system, as
well as the symmetry (\ref{Tsymmetry}) and the reversing
symmetries (\ref{Rsymmetry}).  This is in contrast with the
Runge-Kutta methods which only preserves the symmetry
(\ref{Tsymmetry}), but not the symplectic structure, nor the
reversing symmetries (\ref{Rsymmetry}).  As a consequence the
Runge-Kutta methods do not preserve the KAM tori caused by
symplecticity or reversibility.

\subsection{Numerical Comparison between RK4, RK78 and SI \label{sec42}}
\begin{figure}[h]
  \begin{center}
    \resizebox{.475\textwidth}{!}{
      \includegraphics{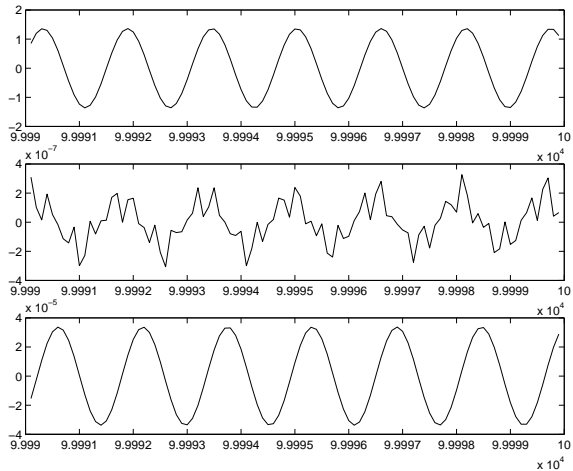}}
  \end{center}
  \caption{
    \label{fig4b}
Plots of the error function $\overline{z}(s_n) - z(s_n)$ against
time.  The upper figure is the result of RK4, the middle figure is
RK78 and the lower figure is of SI.  The time of integration is
$10^5$ with a time step for RK4 and SI of $0.025$.}
\end{figure}

We start by comparing the three numerical methods, i.e. RK4, RK78,
and SI.  We choose the $4:1$-resonance, which is the most
prominent higher order resonance, as a test problem.  We fix the
value of the energy ($H_2$) to be $5$ and take $\varepsilon =
0.05$. Starting at the initial condition $z(0)=0, \varphi(0) = 0,
p_z = \sqrt{5/2},$ and $p_\varphi(0)=0$, we know that the exact
solution we are approximating is given by $(\sqrt{5/2}\;
\sin(4t),0, \sqrt{5/2} \;\cos(4t),0)$. We integrate the equations
of motion up to $t=10^5$ seconds and keep the result of the last
$10$ seconds to have time series $\overline{z}(t_n)$ and
$\overline{p_z}(t_n)$ produced by each integrator. Then we define
a sequence $s_n = 99990 + 5n/100, n = 0,1, \ldots, 200$.  Using an
interpolation method, for each of the time series we calculate the
numerical $\overline{z}(s_n)$. In figure \ref{fig4b} we plot the
error function $\overline{z}(s_n) - z(s_n)$ for each integrator.

The plots in Figure in \ref{fig4b} clearly indicate the
superiority of RK78 compared with the other methods (due to the
higher order method). The error generated by RK78 is of order
$10^{-7}$ for an integration time of $10^5$ seconds. The minimum
time step taken by RK78 is $0.0228$ and the maximum is $0.0238$.
The error generated by SI on the other hand, is of order
$10^{-5}$.  The CPU time of RK78 during this integration is
$667.75$ seconds.  SI completes the computation after $446.72$
seconds while RK4 only needs $149.83$ seconds.

\begin{figure}
  \begin{center}
    \resizebox{.45\textwidth}{!}{
      \includegraphics{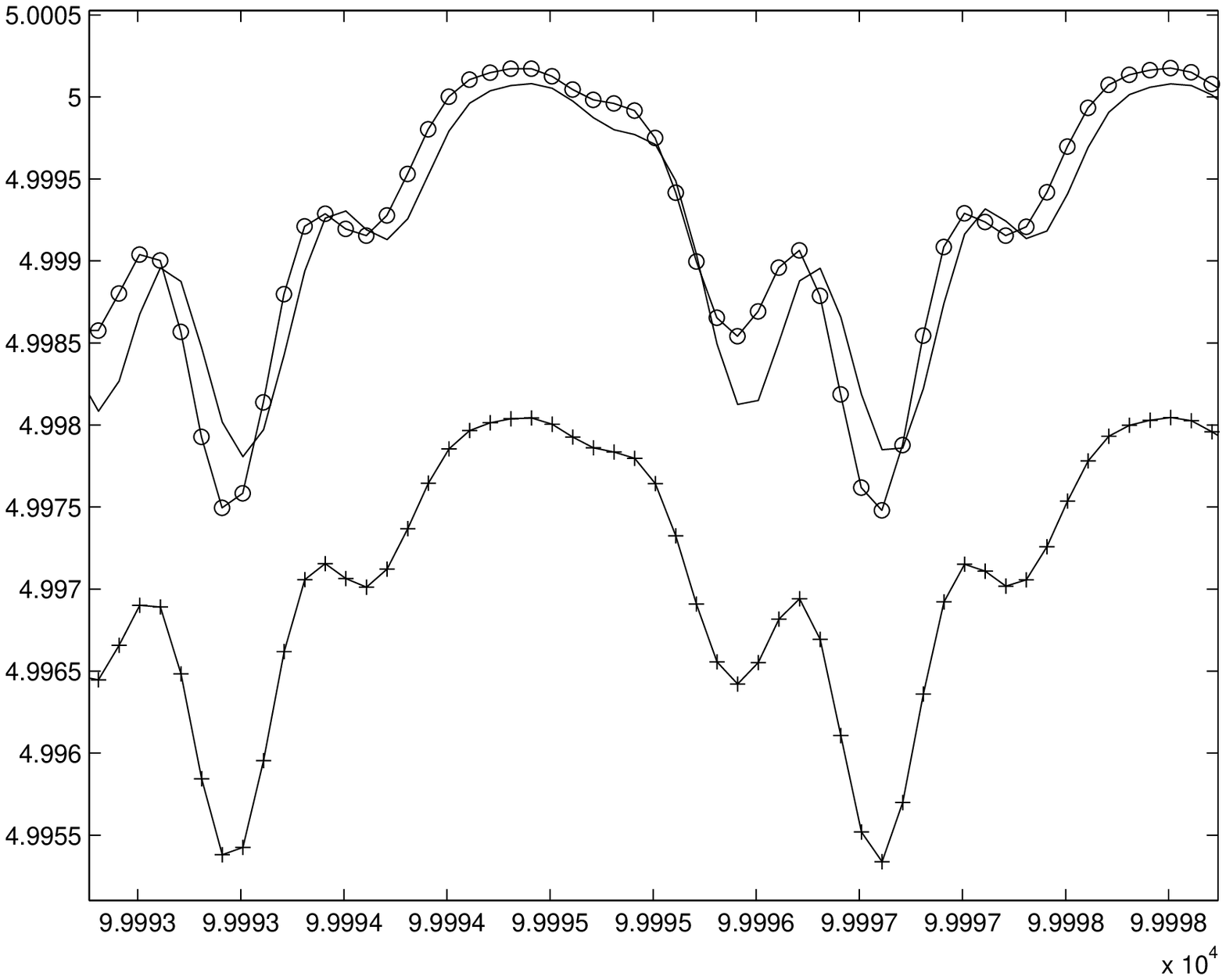}}
     \resizebox{.45\textwidth}{!}{
      \includegraphics{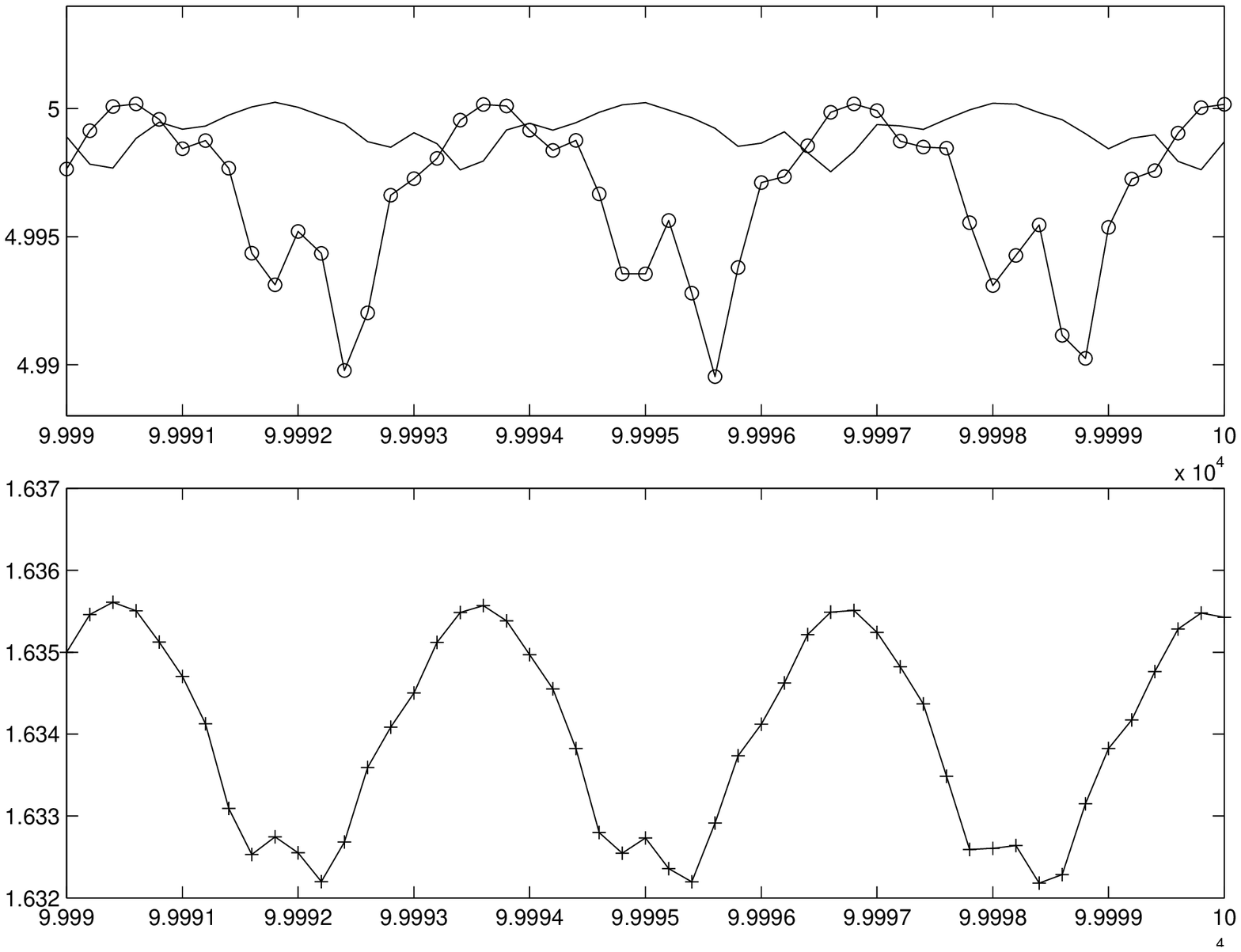}
    }
  \end{center}
  \caption{
    \label{fig1}
Plots of the energy against time.  The solid line represents the
results from SI.  The line with '+' represents the results from
RK4 and the line with '$\circ$' represents the results from RK78.
On the left hand plot, we show the results of all three methods
with the time step $0.01$. The time step in the right hand plots
is $0.05$.  The results from RK4 are plotted seperately since the
energy has decreased significantly compared to the other two
methods.   }
\end{figure}

We will now measure how well these integrators preserve energy. We
start integrating from an initial condition $z(0) = 0$,
$\varphi(0) = 1.55$, $p_\varphi(0) = 0$ and $p_z(0)$ is determined
from $H^3 = 5$ (in other word we integrate on the energy manifold
$H = 5 + O(\varepsilon)$).  The small parameter is $\varepsilon =
0.05$ and we integrate for $t=10^5$ seconds.

For RK78, the integration takes $667.42$ second of CPU time. For
RK4 and SI we used the same time step, that is $10^{-2}$. RK4
takes $377.35$ seconds while SI takes $807.01$ seconds of CPU
time. It is clear that SI, for this size of time step, is
inefficient with regard to CPU time. This is due to the fact that
to construct a higher order method we have to compose the flow
several times. We plot the results of the last $10$ seconds of the
integrations in Figure \ref{fig1}.  We note that in these $10$
seconds, the largest time step used by RK78 is 0.02421\ldots
 while the smallest is 0.02310\ldots.     It is clear
from this, that even though the CPU time of RK4 is very good, the
result in the sense of conservation of energy is rather poor
relative to the other methods.

We increase the time step to $0.05$ and integrate the equations of
motion starting at the same initial condition and for the same
time. The CPU time of SI is now $149.74$ while for the RK4 it is
$76.07$.  Again, in Figure \ref{fig1} (the right hand plots) we
plot the energy against time.  A significant difference between
RK4 and SI then appears in the energy plots. The results of
symplectic integration are still good compared with the higher
order method RK78.  On the other hand, the results from RK4 are
far below the other two.


\subsection{Computation of the Size of the Resonance Domain}
\begin{figure}[h]
  \begin{center}
    \resizebox{10cm}{!}{
      \includegraphics{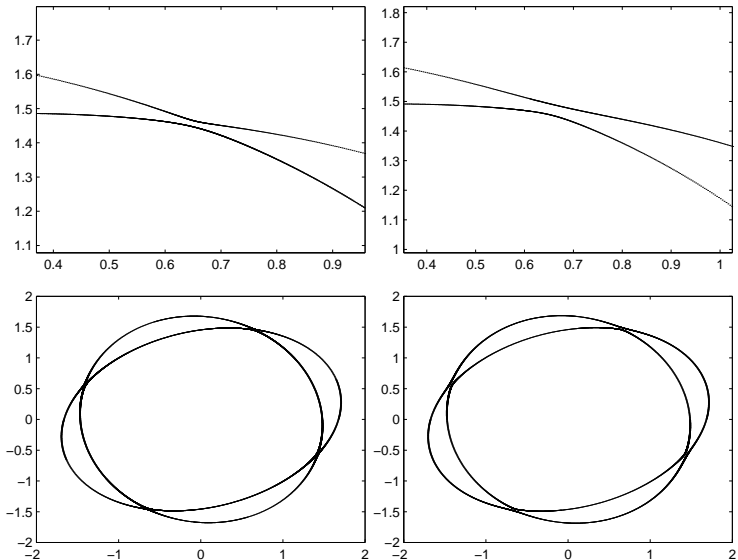}
    }
  \end{center}
  \caption{
    \label{fig3}
Resonance domain for the $4:1$-resonance. The plots on the left
are the results from SI while the right hand plots are the results
from RK78.  The vertical axis is the $p_\varphi$ axis and the
horizontal axis is $\varphi$. The time step is $0.05$ and
$\varepsilon=0.05$. In the top figures, we blow up a part of the
pictures underneath.}
\end{figure}

Finally, we calculate the resonance domain for some of the most
prominent higher order resonances for the elastic pendulum. In
Figure \ref{fig3} we give an example of the resonance domain for
the $4:1$ resonance. We note that RK4 fails to produce the
section.
On the other hand, the results from SI are still accurate.  We
compare the results from SI and RK78 in Figure \ref{fig3}.  After
$5\times 10^3$ seconds, one loop in the plot is completed. For
that time of integration, RK78 takes $34.92$ seconds of CPU time,
while SI takes only $16.35$ seconds.  This is very useful since to
calculate for smaller values of $\varepsilon$ and higher resonance
cases, the integration time is a lot longer which makes it
impractical to use RK78.

In Table \ref{tab1} we list the four most prominent higher order
resonances for the elastic pendulum.  This table is adopted from
\cite{tv} where the authors list six of them.

\begin{table}[h]
\begin{center}
\begin{tabular}{|c|c|c|c|c|}\hline
Resonance & Resonant& Analytic    & Numerical & Error     \\
          &part     & $\log_\varepsilon(d_\varepsilon)$     &  $\log_\varepsilon(d_\varepsilon)$  &   \\ \hline
$4:1$ &  $H_5$ & $1/2$   & $0.5091568$ & 0.004 \\ $6:1$ & $H_7$  &
$3/2 $ & $1.5079998$ & 0.05 \\ $4:3$ &  $H_7$ & $3/2 $   &
$1.4478968$  & 0.09
\\$3:1$ & $H_8$ & $2 $  & $2.0898136$ &  0.35\\
  \hline
\end{tabular}
\caption{\label{tab1}  Comparison between the analytic estimate
and the numerical computation of the size of the resonance domain
of four of the most prominent higher order resonances of the
elastic pendulum.  The second column of this table indicates the
part of the expanded Hamiltonian in which the lowest order
resonant terms are found. }
\end{center}
\end{table}

\begin{figure}[h]
  \begin{center}
    \resizebox{8cm}{!}{
      \includegraphics{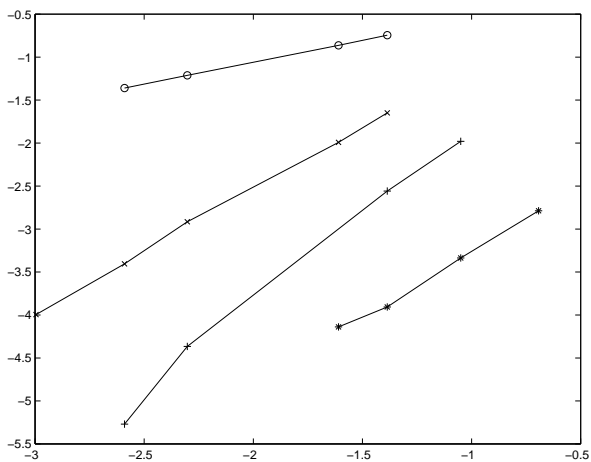}
    }
  \end{center}
  \caption{
    \label{fig6}
Plots of $\log(d_\varepsilon)$ against $\log(\varepsilon)$ for
various resonances.  The $4:1$-resonance is plotted using
'$-\circ$', the $3:1$-resonance is using '$-+$', the
$4:3$-resonance is using '$-\times$' and the $6:1$ resonance is
using '$-\ast$'.}
\end{figure}

The numerical size of the domain in table \ref{tab1} is computed
as follows. We first draw several orbits of the Poincar\'e maps
$\mathcal{P}$. Using a twist map argument, we can locate the
resonance domain.  By adjusting the initial condition manually, we
then approximate the heteroclinic cycle of $\mathcal{P}$.  See
figure \ref{fig3} for illustration.  Using interpolation we
construct the function $r_o(\theta)$ which represent the distance
of a point in the outer cycle to the origin and $\theta$ is the
angle with respect to the positive horizontal axis.  We do the
same for the inner cycle  and then calculate $\max_\theta
|r_o(\theta) - r_i(\theta)|$.  The higher the resonance is, the
more difficult to compute the size of the domain in this way.

For resonances with very high order, manually approximating the
heteroclinic cycles would become impractical, and one could do the
following. First we have to calculate the location of the fixed
points of the iterated Poincar\'e maps numerically. Then we can
construct approximations of the stable and unstable manifolds of
one of the saddle points. By shooting to the next saddle point, we
can make corrections to the approximate stable and unstable
manifold of the fixed point.

\section{Discussion}
In this section we summarize the previous sections.  First the
performance of the integrators is summarized in table \ref{tab2}.

\begin{table}[h]
\begin{center}
\begin{tabular}{||c|c||c|c|c||}\cline{3-5}
\multicolumn{2}{c||}{} & \multicolumn{3}{c||}{Integrators} \\
\cline{3-5}
\multicolumn{2}{c||}{}     & RK4 & RK78 & SI  \\ \hline \hline   &
CPU time {\scriptsize ($\Delta t \approx 0.025, t = 10^5$ sec.) }
& $149.83$ sec.   & $667.75$ sec.
 & $446.72$ sec.\\ \cline{2-5}
The $4:1$  & Preservation of $H$ {\scriptsize ($t=10^5$ sec.)} &
Poor & Good & Good  \\ \cline{2-5}
 resonance     & Orbital Quality &
Poor  & Very good  &  Good  \\ \cline{2-5}  & Section Quality &
--- & Good & Good  \\ \cline{1-5} The $6:1$ & Orbital Quality  &
Poor & Good  &  Good
\\ \cline{2-5} resonance & Section Quality & --- & Good & Good
\\ \cline{1-5} The $4:3$ & Orbital Quality  & Poor & Good  & Good
\\ \cline{2-5} resonance & Section Quality  & --- & --- & Good
\\ \hline
The $3:1$ & Orbital Quality  & Poor  & Poor  &  Good  \\
\cline{2-5} resonance & Section Quality  & --- & --- & Good  \\
\hline\hline
\end{tabular}
\caption{Summary of the performance of the integrators.  A bar ---
indicates that it is not feasible to obtain a surface of section
for this resonance using this integrator.}

\label{tab2}
\end{center}
\end{table}
As indicated in table \ref{tab2}, for the $4:3$ and the $3:1$
resonances, the higher order Runge-Kutta method fails to produce
the section. This is caused by the dissipation term, artificially
introduced by this numerical method, which after a long time of
integration starts to be more significant.  On the other hand, we
conclude that the results of our symplectic integrator are
reliable.  This conclusion is also supported by the numerical
calculations of the size of the resonance domain (listed in Table
\ref{tab2}).

In order to force the higher order Runge-Kutta method to be able
to produce the section, one could also do the following.  Keeping
in mind that RK78 has automatic step size control based on the
smoothness of the vector field, one could manually set the maximum
time step for RK78 to be smaller than $0.02310$.  This would make
the integration times extremely long however.

We should remark that in this paper we have made a number of
simplifications.   One is that we have not used the original
Hamiltonian.  The truncated Taylor expansion of (\ref{transform})
is polynomial.  Somehow this may have a smoothing effect on the
Hamiltonian system.  It would be interesting to see the effect of
this simplification on the dynamics of the full system. Another
simplification is that, instead of choosing our initial conditions
in the energy manifold $H=C$, we are choosing them in $H^3 = C$.
By using the full Hamiltonian instead of the truncated Taylor
expansion of the Hamiltonian, it would become easy to choose the
initial conditions in the original energy manifold. Nevertheless,
since in this paper we always start in the section $\Sigma$, we
know that we are actually approximating the original energy
manifold up to order $\varepsilon^2$.

We also have not used the presence of the small parameter
$\varepsilon$ in the system.  As noted in \cite{Mc}, it may be
possible to improve our symplectic integrator using this small
parameter.  Still related to this small parameter, one also might
ask whether it would be possible to go to even smaller values of
$\varepsilon$.  In this paper we took $e^{-3} < \varepsilon <
e^{-0.5}$.  As noted in the previous section, the method that we
apply in this paper can not be used for computing the size of the
resonance domain for very high order resonances. This is due to
the fact that the resonance domain then becomes exceedingly small.
This is more or less the same difficulty we might encounter if we
decrease the value of $\varepsilon$.  

Another interesting posibility is to numerically follow the resonance manifold as the energy increases.  As noted in the introduction, this is numerically difficult problem.  Since this symplectic integration method offers a cheap and accurate way of producing the resonance domain, it might be posible to numerically study the bifurcation of the resonance manifold as the energy increases.  Again, we note that to do so we would have to use the full Hamiltonian.

\vskip 1cm

{\noindent {\bf Acknowledgements}

J.M. Tuwankotta thanks the School of Mathematical and Statistical
Sciences, La Trobe University, Australia for their hospitality
when he was visiting the university.  Thanks to David McLaren of
La Trobe University, and Ferdinand Verhulst, Menno Verbeek and
Michiel Hochstenbach of Universiteit Utrecht, the Netherlands for
their support and help during the execution of this research. Many
thanks also to Santi Goenarso for every support she has given.

We are grateful to the Nederlandse Organisatie voor
Wetenschappelijk Onderzoek (NWO) and to the Australian Research
Council (ARC) for financial support.}

\end{document}